\documentclass[fleqn, a4paper]{article}
\usepackage[hyphens]{url}
\usepackage[bookmarks=true]{hyperref}
\usepackage{natbib,mathtools,graphicx,enumitem,txfonts,subcaption,csquotes,doi,booktabs,cleveref}
\usepackage[vmargin=0.87in, hmargin=0.9in]{geometry}
\usepackage[strings]{underscore}
\graphicspath{{Figures/}{./}}
\renewcommand{\(}{\left(}
\renewcommand{\)}{\right)}
\newcommand{\W}{\mathcal{W}}
\newcommand{\M}{\mathcal{M}}
\newcommand{\ra}[1]{\renewcommand{\arraystretch}{#1}}
\let\ab\allowbreak

\title{Coalitions \& Voting Power in the Greek Parliament of 2012: A Case-Study}
\author{Constandina Koki\\
AUEB, Greece\\
kokiconst@aueb.gr \and Stefanos Leonardos\\ SUTD, Singapore\\stefanos_leonardos@sutd.edu.sg}

\begin{document}
\maketitle

\begin{abstract}
We revisit the May and June 2012 Greek Parliamentary elections and the December 2014 Presidential election that was held by the June-elected Parliament. The three voting instances provide a political field experiment for the application of power indices and their interpretation in context. We model the Greek Parliament as a weighted majority game and assess voting power with the Shapley-Shubik, Holler and when relevant, Coleman's indices. Also, based on the actual events, we establish connections between parties and evaluate the Myerson index. We focus on the influence of institutional rules on the distribution of power among the elected political parties and add an alternative input to the ongoing political debate about the reform of both the Parliamentary and Presidential electoral system in Greece. Additionally, our findings contribute to the understanding of the coalition formation process in the particular context and provide empirical evidence on the performance of non-selective indices in parliamentary multi-party settings which can be used for comparison by similar case-studies in the future.
\end{abstract}
\textbf{Keywords:} Decision Making, Weighted Voting Games, Power Indices, Greece, 2012 Elections\\
\textbf{JEL Classification:} D7:D72, C7:C71.

\section{Introduction}\label{introduction}
The political events that took place in Greece between 2012 and 2015, see \cite{Va13, Te14}, attracted widespread international attention. Having signed two loan agreements -- Memorandums of Understanding (MoUs) -- that carried harsh austerity measures to avoid bailout, Greece was facing severe economic problems. The public was divided between supporters and opponents of the MoUs in a new cleavage that transcended the traditional left-right division. In this environment of political instability, uncertainty about the future and extended social protests, Greece went to early elections in May 2012. In the elections, $7$ parties gained parliamentary representation. These were -- from left to right -- KKE, SYRIZA, DIMAR, PASOK, ND, ANEL and GD, \cite{Di13}. With agendas of various intensity, KKE, SYRIZA, ANEL and GD were against the implementation of the MoUs, whereas ND and PASOK were in favor. DIMAR held a neutral, moderate stance, placing itself as a stabilizing political actor. ND won the elections by a short margin to SYRIZA. However, due to fragmentation of the electorate's preferences among the various parties within the pro/anti MoU camps, the elections could not produce a majority government. After the elections, ND quickly aligned with PASOK, but negotiations with DIMAR to form a  government led to a deadlock and new elections were called for June 2012. \par
Under more polarized conditions, ND raised its vote share by $10\%$ and again won the elections by a short but clear margin from SYRIZA. This time and under both domestic and international pressure for a stable outcome, the negotiations were conclusive and led to an agreement between ND, PASOK and DIMAR. However, the coalitional government proved fragile and short-lived. Soon thereafter, and under constant friction between its members, DIMAR abandoned the coalition, posing ND and PASOK with a formidable challenge: to stay in power, the two remaining governing partners had to ensure a $60\%$ enhanced majority in the Parliament to elect the President of the Republic, \cite{Ko15}. The opposition managed to block the election of President, and in December 2014, the Parliament was dissolved. Elections were called for January 2015 leading to the formation of the currently (as of 2018) incumbent governing coalition between SYRIZA and ANEL, \cite{Ts16}. \par
In this paper, we employ quantitative tools for the measurement of voting power, see \cite{Fe98,Na19}, to re-analyze the political events of the 2012-14 period from a mathematical point of view. The motivation for our study is twofold. From a political perspective, these events are still relevant at both domestic and international level. Along the lasting impact on European politics, the instability caused by these 3 elections has triggered a still ongoing, political debate about the amendment of both the Parliamentary and the Presidential electoral rules in Greece, \cite{ek16,na18,Fo18}. From a methodological perspective, these elections constitute a rare field-experiment for the application and comparison of power indices. The two Parliamentary elections were held in a very short period (within 6 weeks) under very similar circumstances. In both elections, the same $7$ parties entered the Parliament and engaged in extensive negotiations to form governing coalitions. While multi-party settings already pose a difficult test for power indices, see \cite{Ho18}, the Greek Parliament of 2012 presents an additional challenge: dictated by the pro vs anti MoU cleavage, that transcended the traditional left-right axis, the connections between these $7$ parties were unusually dense and rather unexpected. Finally, as a dichotomous -- yes/no -- voting procedure, the Presidential election provides an additional opportunity for the evaluation and study of power indices with the same parties but at the enhanced $60\%$ quota.

\subsection{Methodology}
By lack of consensus on the \emph{right index}, see \cite{Ho13,Ho14p,Ho01}, we calculate the most common indices and interpret them with different criteria, see \cite{La95, Le02} and \cite{Al11} among others. We represent three different instances -- May and June 2012 and December 2014 -- of the Greek Parliament as \emph{weighted majority games} and evaluate the Shapley-Shubik, Penrose-Banzhaf, Deegan-Packel and Public Good (Holler) indices for the $50\%$ majority and additionally the Coleman's indices to initiate and to prevent action for the $60\%$ majority, see also \cite{Le02,Al08,AMe13,Zu12}. Based on the parties' political agendas and public statements about potential coalitions, we establish a \emph{communication structure} or set of connections between them and evaluate the Myerson index, \cite{My77}. Calculations have been performed with the freely available \href{https://alex.uniupo.it/software}{AL.EX4} and \href{http://homepages.warwick.ac.uk/~ecaae/ipdirect.html}{ipdirect} software programs.\par
Due to the ordinal equivalence of the Shapley-Shubik and Banzhaf indices in weighted majority games, \cite{To87, Fr10}, the selection of a specific index does not affect the power-hierarchy among parties. Our calculations confirmed this property which allowed us to drop the Banzhaf index (absolute and normalized) and focus only on the Shapley-Shubik index. We further exploited this feature and focus on their induced rankings instead of their specific values when measuring \enquote{party-influence} rather than \enquote{payoffs}, see \cite{Fe98,Fr12}. Similarly, our calculations led to equivalent results between the Deegan-Packel and Holler indices, from which we kept only the latter.

\subsection{Summary of results}
 In the May 2012 elections, the Holler Index (PGI) attributes a disproportionally high power to smaller parties. This indicates an uneven distribution of power away from the major parties that may have hurdled the process of government formation. Remarkably, while SYRIZA is second according to both the Shapley-Shubik Index (SSI) and the actual vote share, it is ranked second to last according to the PGI. Compared to the non-selective indices, the Myerson index utilizes DIMAR's dense connections and accurately captures its influence in the negotiations by attributing to it its second highest value of $22\%$ despite DIMAR's $6.33\%$ parliamentary seat share. \par
 In the June 2012 elections, SYRIZA's SSI value is equal to $12\%$ despite its $23.67\%$ share of seats. This ranks SYRIZA equally with PASOK, indicating that both parties wield the same power to influence the negotiations despite a $14\%$ difference in their vote shares. When it comes to minimal winning coalitions, the PGI ranks SYRIZA second to last. The low a-priori power of SYRIZA may be attributed to the current electoral rule of reinforced proportionality and hence, may provide an argument in favor of its revision as Greece transitions from a bipartisan to a multi-party system, \cite{ek16}. Equally interesting are the findings by the informative Myerson index. Being not monotonic in voting weights, and despite a decrease in PASOK's share, it now correctly identifies ND and PASOK as the two main political actors due to their ability to form a government without a third partner. In view of ND's and PASOK's connection and despite a considerable increase in its seat share, SYRIZA's Myerson value decreases. Similarly, DIMAR is shown to forfeit the advantage of being a crucial player in the negotiations. \par
Finally, the evaluation of Coleman's index of a party to prevent action for the Presidential election of 2014 at the $60\%$ quota reveals an inflated power of the opposition to form blocking coalitions. Given the increasing representation of extreme, anti-systemic parties in the Parliament and their \emph{de facto} alignment against any action, this supports the current debate about the necessity to carefully rethink the Presidential electoral rule as part of a broader Constitutional revision, \cite{na18,he18}.

\subsection{Outline}
The rest of the paper is structured as follows. In \Cref{preliminaries}, we provide the mathematical definitions of cooperative games, weighted voting games and power indices and discuss the Greek historic and political context, before and during the twin elections of 2012. \Cref{may,june} contain the evaluation of power indices for the May and June 2012 elections and their interpretation in the specific political context. In \Cref{presidential}, we study the December 2014, Presidential election. We summarize our conclusions in \Cref{conclusions} along with suggestions for future research.

\section{Preliminaries}\label{preliminaries}
To make the paper self-contained, \Cref{cooperative} summarizes the notions of weighted voting games and power indices, using standard notation as in \cite{Fe98,Ta99,Le02} and \cite{Na19}. Since our focus is on the interpretation of power indices in the specific context of the Greek elections, a reader may skip to \Cref{context} without significantly compromising the understanding of the paper.

\subsection{Mathematical definitions}\label{cooperative}
We consider a finite set of \emph{voters or parties}, $N=\{1,2,\dots,n\}$. Collections $S\subseteq N$ of voters are called \emph{coalitions}. A \emph{simple voting} game $G = \(N,v\)$ on the set $N$ is defined by a mapping $v: S\to \mathbb\{0,1\}$, which is called the \emph{characteristic function} of the game $G$. A coalition $S\subseteq N$ is said to be \emph{winning} if $v\(S\)=1$, and \emph{losing} if $v\(S\)=0$. The empty set $\emptyset$ is a losing coalition, i.e., $v\(\emptyset\)=0$, and the coalition of all players is winning, i.e., $v\(N\)=1$. Also, we assume that $v$ is \emph{monotonic}, i.e., $v\(S\)\le v\(T\)$ whenever $S\subset T$. A simple game $v$ is completely determined by the set of \emph{winning} coalitions $\W\:=\{S\subseteq N: v\(S\)=1\}$ or due to monotonicity, by the set of \emph{minimal} winning coalitions $\M:=\{S\in \W: T\subset S \implies T \notin \W\}$. Accordingly, for each player $i\in N$, the sets of winning and minimal winning coalitions that contain player $i$ are denoted by $\W_i:=\{S\in \W: i\in S\}$ and $\M_i:=\{S\in \M: i\in S\}$. 

\subsubsection{Weighted voting games}\label{weighted}
\cite{Ho01} argue that most parliaments of Western Democracy in fact operate and hence may be studied as weighted voting systems. A voting body can be thought of as a \emph{weighted voting} game, $G=[q; w_1,\dots,w_n]$, where $G$ is a simple game, $w = \(w_1, \dots, w_n\) \in \mathbb R_+^n$ a vector of nonnegative real \emph{weights}, and $q\in \mathbb R_+$ a nonnegative \emph{quota} with the following decision rule: a coalition $S\subseteq N$ is winning, $v\(S\)=1$, if and only if its \emph{total weight}, $w\(S\):=\sum_{i\in S}w_i$, meets or exceeds the quota, i.e., 
\[v\(S\)=\begin{cases}1, & \text{if } w\(S\)\ge q,\\ 0, & \text{otherwise}.\end{cases}\]
If $\frac12\sum_{i\in N} w_i <q \le \sum_{i\in N}w_i$, then $G$ is also called a \emph{weighted majority} game. While weighted voting games are simple games not every simple game can be represented as a weighted voting game, see \cite{Hof18}. A player $i\in N$ is called \emph{pivotal} or \emph{crucial} for a coalition $S\subseteq N\setminus\{i\}$ if $v\(S\cup\{i\}\)=1$ and $v\(S\)=0$, i.e., if $w\(S\) < q$ and $w_i+w\(S\)\ge q$. For each $i \in N$, let $\eta_i$ denote the set of all coalitions $S$ for which player $i$ is crucial, i.e., $\eta_i:=\{S\in \W: S\setminus\{i\}\notin \W\}$. Every coalition $S\in\eta_i$ is called a \emph{swing} for player $i$. A player $i$ is called a \emph{null} player if she does not contribute to any coalition, i.e., if $v\(S\cup \{i\}\)=v\(S\)$ for all $S\subseteq N\setminus\{i\}$. To exclude the possibility of having two contradictory decisions, i.e., two simultaneously winning coalitions, we assume that the complement $N\setminus\{S\}$ of a winning coalition $S$ is losing, i.e., that $v\(N\setminus\{S\}\)=0$ whenever $v\(S\)=1$. However, it is possible that for some coalition $S$, both $S$ and $N\setminus\{S\}$ are losing. In this case, $S$ is called a \emph{blocking} coalition. This may occur for instance when a supermajority $q>50\%$ is required for some decision, see also \Cref{presidential} for an example.

\subsubsection{Classic power indices}\label{poweri}
A \emph{power index} $\psi$ assigns a non-negative real number to every voter in a weighted voting game $G$. To simplify notation, we will write $\psi_i$ instead of $\psi_i\(G\)$ to denote the value that power index $\psi$ assigns to voter $i\in N$ in game $G$. The non-negative real number $\psi_i$ is interpreted as the \emph{power} of the corresponding player $i \in N$ in game $G$ and can be generally thought of as numerical estimate of the \emph{a priori} influence or capacity of being decisive of each player/party in a weighted voting game, \cite{Be15}. The \emph{Shapley-Shubik index} (SSI), \cite{Sh54}, assigns to every player $i\in N$ the real number 
\[\phi_i:=\sum_{S\subseteq N}\frac{s!\cdot\(n-s-1\)!}{n!}\cdot \left[v\(S\cup\{i\}\)-v\(S\)\right]\]
where $s$ denotes the cardinality of set $S$, i.e. $s:=|S|$. 
It is interpreted as a measure of \emph{power as prize} or payoff (P-power), \cite{Fe98}, and is meant to be a voter's (party's) expected payoff -- as a percentage -- from entering a winning coalition, \cite{Du79,Fe98}.The \emph{Holler or Public Good Index} (PGI), \cite{Ho82,Ho83}, assigns to every player $i\in N$ the real number 
\[\delta_i:=\frac{\left|\M_i\right|}{\sum_{j\in N}\left|\M_j\right|}\]
which is equal to the total number of minimal winning coalitions containing player $i$, divided by the sum of all minimal coalitions for all players $j\in N$. The properties of the PGI index are discussed in\cite{Lo07}.\par
\cite{Co71} proposes two indices to measure the power of each player. The power of a member $i$ \emph{to initiate action} $CI_i$ measures the member's potential to swing a coalition from losing to winning and is defined as the number of $i$'s swings relative to the total number of losing coalitions, $CI_i=\eta_i/\(2^n-|\W|\)$. The power of a member $i$ \emph{to prevent action} $CP_i$ measures the member's ability to block a decision by means of a swing and is defined as the proportion of winning coalitions that are also swings for player $i$,  $CP_i=\eta_i/|\W|$. The distinction of power to initiate and power to prevent an action only matters for bodies which employ a supermajority $q$, strictly greater than $50\%$, \cite{Le02}. Since Coleman's indices are equal and equivalent to Banzhaf index at the $50\%$ quota, we use them only in the context of the Presidential election at the $q=60\%$ quota, see \Cref{presidential}.

\subsubsection{Connections between parties \& Myerson index}
The SSI and PGI index are non-selective, i.e., deem all possible coalitions as possible and hence provide an \emph{a-priori} measurement of voting power. At interim situations, one may exploit existing information about established connections between parties to evaluate the \emph{Myerson index}, \cite{My77}. Formally, given a simple game $G=\(N,v\)$, a \emph{communication structure} on $G$ is defined as a set $E$ of \emph{connections or links} between parties $m,n\in N$. Let $\mathcal F:=\{S\subseteq N: \text{the members of $S$ are connected according to communication structure $E$}\}$ denote the set of all \emph{feasible coalitions} given the communication structure $E$. For any $S\subseteq N$ let $C\(S\)$ denote the set of all maximal feasible subsets $T\subseteq S$. Then, the $E$-restricted game $G_E=\(N,v_E\)$ is defined by $v_E:=\sum_{T\in C\(S\)} v\(T\)$. Then, the value $\mu_i$ of the \emph{Myerson index} for party $i \in N$ in $G$ is equal to the Shapley-Shubik index computed on game defined by the characteristic function $v_E$, i.e., $\mu_i:=\phi_i\(G_E\)$. If all parties are connected, then the Myerson Index reduces to the Shapley-Shubik index. This formulation captures the idea that even if two parties are not connected -- do not share a link between themselves -- they may still effectively cooperate if they both have a connection with the same mutual third party, \cite{My77,Fe02}.

\subsection{Political context}\label{context}
We start with a brief description of the historical and socio-economic background in which the elections took place, based on \cite{Di13}, \cite{Te14} and \cite{Va13}. More details can be found in \cite{Ne12,Ko15,Ma16b} and \cite{Ts16}.

\subsubsection{Historical background}\label{history}
Since the restoration of democracy in 1974 and up to 2009, the Greek party system has been a typical example of two-partyism with the combined percentage of the two dominant parties, the center-left PASOK and center-right ND, often exceeding $85\%$ of the total vote. A disproportional electoral system which assigns a bonus of seats to the first party has enhanced stability in this bipartisan political environment and has aided the formation of single-parity majority governments. As such, Greek politics lack a history and culture of consensus and communication, \cite{Va13,Di15} and coalitions have been rare and short-lived, \cite{Ne12}.

\subsubsection{Parliamentary \& Presidential electoral rules}\label{rule}
Members of Parliament (MPs) are elected for 4-year terms in the 300-seat legislature (Vouli ton Ellinon). The Parliamentary electoral system in effect during both the 6 May and 12 June elections was a form of semi or \emph{reinforced} proportional representation with a majority bonus: the party with the plurality of votes cast was awarded an extra 50 seats out of a total of 300 parliamentary seats. The remaining 250 seats were divided proportionally according to each party's total valid vote percentage. Small parties had to reach a threshold of $3\%$ to be represented in Parliament. The premium of $50$ parliamentary seats was received primarily at the expense of the second-place party and the parties not entering the parliament \cite{Ka07}.\par
The President of the Republic is elected by the Parliament for 5-year terms. This is usually done via a binary (yes/no) voting process, in which the governing party or coalition proposes a candidate who according to the Greek Constitution (Article 32) needs to reach $60\%$ of the votes, or equivalently 180 out of 300 MP votes, to be elected\footnote{The actual process is a slightly more complex. Here we restrict attention to the high level description that is necessary for our purpose.}. If such a majority is not achieved, the Parliament is dissolved and early elections are called, \cite{Ko15}. Then, the new Parliament can elect the President with a simple majority of 151 out of 300 MPs. Based on the above and despite the ceremonial nature of this post, the Greek Constitution effectively gives a veto to a coalition of opposition parties if they can muster at least 121 (from a total of 300) votes, \cite{Ts16}. 

\subsubsection{Party connections and their political positions in 2012}\label{system}
The twin 2012 elections signaled a radical change in the Greek political scene. The electoral results pointed to a transition from the stable two-party system to a uniform distribution of power among several parties. The traditional left-right cleavage that had served as a safe predictor of the Greek public's voting preferences gave its place to a new demarcation line between Euro advocates and skeptics and between friends and foes of the MoUs, \cite{Di15}. The main positions of the $7$ elected parties are briefly summarized in the last two columns of \Cref{connect2012}. The parties are presented according to their position in the left-right ideological axis (KKE left to GD far right).\par
The pro-MoU camp was represented by the traditional parties of ND and PASOK. DIMAR adopted a more moderate stance concerning the necessity of the MoUs, but proclaimed that Greece's prospect in the EU was non-negotiable. The anti-MoU camp was more popular, however, its power was fragmented and distributed among many parties -- many of which did not enter the parliament -- with different ideology and agenda. Its main delegate was SYRIZA in the left followed by ANEL in the right. Their \enquote{conditionally pro Euro-zone} entries capture their rhetoric that they had a pro-Euro disposition but that they would consider exiting the Euro-zone in case Greece's European partners and lenders were not willing to reconsider the debt agreements. 

\setlength{\tabcolsep}{5.6pt}
\begin{table}[!htb]
\centering	
\begin{tabular}{@{}lccccccccll@{}}
\cmidrule{1-8}\cmidrule{10-11}
&\multicolumn{7}{c}{\bf Connectivity Graph}& & \multicolumn{2}{c}{\bf Main Positions}\\[0.12cm]
& \bf KKE & \bf SYRIZA &\bf DIMAR & \bf PASOK & \bf ND &\bf  ANEL & \bf GD && \bf Memorandum & \bf Euro-zone\\\cmidrule{1-8}\cmidrule{10-11}
\bf KKE  & $\bullet$& $\bullet$&&&&&&& anti & anti \\
\bf SYRIZA &$\bullet$& $\bullet$&$\bullet$ & &&$\bullet$& && anti & conditionally pro\\
\bf DIMAR && $\bullet$ & $\bullet$ & $\bullet$&$\bullet$&& && moderate & pro\\
\bf PASOK &&&$\bullet$&$\bullet$&$\bullet$&& && pro & pro\\
\bf ND && &$\bullet$&$\bullet$&$\bullet$&& && pro & pro\\
\bf ANEL && $\bullet$ & && & $\bullet$& && anti & conditionally pro\\
\bf GD &&& & & &&$\bullet$ && anti & anti\\\cmidrule{1-8}\cmidrule{10-11}
\end{tabular}
\caption{Connectivity graph and main positions (Memorandum and Euro-zone columns) of the parties in the May 06 \& June 12, 2012 negotiations.}
\label{connect2012}
\end{table}
The Connectivity Graph in the first part of \Cref{connect2012} describes connections between parties based on their public statements about potential coalitions prior and between the twin elections. Due to the pro versus anti MoU cleavage, the connections are unusually dense and transcend the traditional left-right axis. In the pro-MoU camp, former foes, ND and PASOK, now stressed the necessity to align forces and form a broad and stable pro-MoU government. DIMAR assumed the role of a stabilizing political factor with moderate stance willing to participate in any pro-EU governing coalition. In the anti-MoU camp and despite their ideological differences, the radical-left SYRIZA and populist-right ANEL advocated the prospect of cooperation on the basis of an anti-MoU government. SYRIZA was also open to discussions with DIMAR and the ideologically adjacent KKE. However, KKE officially rejected any prospect of cooperation with SYRIZA. Given KKE's hard stance, SYRIZA directly addressed KKE's voters and indeed poles indicated that almost $50\%$ of KKE's voters in May, switched to SYRIZA in June. Hence, the indicated connection between KKE and SYRIZA should be understood at voters' level. On the right extreme, GD excluded their participation in any potential government.

\section{The May 06, 2012 Parliamentary election}\label{may}
\begin{figure}[!htb]
\begin{minipage}[b]{0.49\textwidth}\centering
\begin{tabular}{@{}llrrr@{}}
\toprule
&&\multicolumn{3}{c}{May 06, 2012} \\
\cmidrule{3-5}
\multicolumn{2}{@{}l}{Political Party} & \% Votes & \# Seats & \% Seats\\
\midrule
ND  && 18.85\% & 108 & 36.00\% \\
SYRIZA && 16.78\% & 52 & 17.33\%\\
PASOK  && 13.18\% & 41 & 13.67\%\\
ANEL  && 10.60\% & 33 & 11.00\% \\
KKE  && 8.48\% & 26 & 8.67\%\\
GD  && 6.97\% & 21 & 7.00\% \\
DIMAR && 6.11\% & 19 & 6.33\% \\
\bottomrule
\end{tabular}
\label[table]{parliamentmay}
\captionof{table}{Votes share, Seats and Seat share.}
\end{minipage}
\hfill
\begin{minipage}[b]{0.49\textwidth}
\centering
\includegraphics[width=\linewidth]{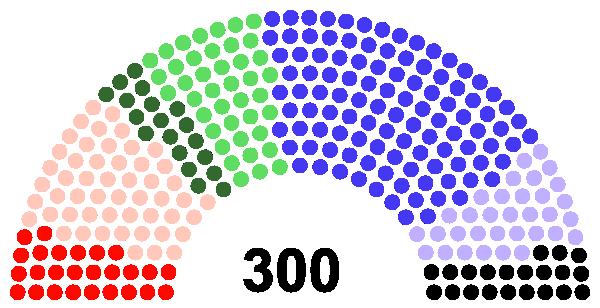}
\captionof{figure}{Parliament (left to right), May 06, 2012.}
\label{parliamentmayfig}
\end{minipage}
\end{figure}

The results of the May 06, 2012 elections are presented in \Cref{parliamentmay} and \Cref{parliamentmayfig}\footnote{Source: \cite{Pa12} and \href{http://www.ypes.gr/el/Elections/NationalElections/}{Greek Ministry of Interior}. To create \Cref{parliamentmayfig}, we used: \href{https://tools.wmflabs.org/parliamentdiagram/parlitest.php}{WmfLabs}.}. The 50 seats bonus has been already assigned to ND, i.e., ND's seats are equal to $58+50=108$. 
Designed to assist the formation of majority governments in a two-party system, the electoral rule could not handle the result. As a consequence, a considerable $20\%$ of the total vote went to parties that did not manage to exceed the $3\%$ threshold and gain parliamentary representation. Coalition formation to produce a government was necessary. ND and PASOK quickly agreed to become partners and in view of SYRIZA's and ANEL's hard stance, DIMAR was the main candidate to complete the government. However, amidst increasing polarization, DIMAR refused to cooperate, the discussions led to a stalemate and new elections were called for June 12, 2012.

\subsection{Measuring a priori power: the classic indices}\label{indicesmay}
\begin{figure}[!htb]
\centering
\includegraphics[]{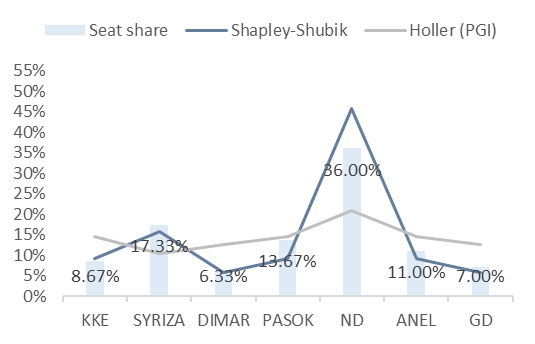}
\caption[]{Parliamentary seats share and main P-Power indices after May 06, 2012 elections.}
\label{powersharemay}
\end{figure}

The Greek Parliament after the May 06, 2012 elections can be described as the weighted voting game  $G= [151; 108, \ab 52, 41, 33, 26, 21, 19]$. The 151 seats correspond to simple majority, $q=50\%$. \Cref{may50} shows the values of the Shapley-Shubik Index (SSI) and Holler or Public Good Index (PGI) which are also depicted graphically -- along with the parliamentary seat share of the parties -- in \Cref{powersharemay}.
\begin{table*}[!htb]
\centering
\ra{1}
\begin{tabular}{@{}llrrrrrrr@{}}\toprule
&& \multicolumn{7}{c}{Political parties: May 06, 2012} \\
\cmidrule{3-9}
Power index & & KKE & SYRIZA & DIMAR & PASOK & \phantom{PS}ND & ANEL & \phantom{PS}GD\\
\midrule
Shapley-Shubik && 0.09 & 0.16 & 0.06 & 0.09 & 0.46 & 0.09 & 0.06\\
Holler && 0.15 & 0.10 & 0.13 & 0.15 & 0.21 & 0.15 & 0.13 \\
\bottomrule
\end{tabular}
\caption{Majority at 50+\% (151 MPs): a priori voting power in May 06, 2012.}
\label{may50}
\end{table*}
Our main findings are the following. First, the SSI largely coincides with the parliamentary seat share of all parties except for the winning party, ND. Interpreted as a measure of P-power, \cite{Fe98}, the SSI shows that ND anticipates an inflated share on the potential spoils from forming a government. ND's SSI is equal to $46\%$ compared to its seat share of $36\%$. Second, the PGI -- which considers only minimum winning coalitions -- attributes a disproportionally high power to the smaller parties. This captures the uneven representation of the vote and distribution of power away from the major parties which can hurdle the process of government formation. Accordingly, if taken into account, it may give an incentive for strategic voting to voters who prioritize stability (parties that are likely to win the elections) over their true party preference. When focusing on the induced rankings of the political parties instead of their specific values, see also \cite{Fr12}, we observe that even though PASOK, ANEL and KKE have different seat shares, they have the same rank and hence the same power to influence the negotiations (I-power). The same applies for GD and DIMAR. Finally, it is remarkable that while SYRIZA is second according to both the SSI and the actual vote share, it is ranked second to last according to the PGI. This may reflect the intuition captured by \cite{Ho14p} that \enquote{it could well be that a larger player is not always welcome to form a winning coalition if a smaller one does the same job}.

\subsubsection{Connections between parties and Myerson index}
Based on the connectivity graph (\Cref{connect2012}), we calculate the Myerson index, which is given in \Cref{myersonmay}.
\begin{table*}[!htb]\centering
\ra{1}
\begin{tabular}{@{}llrrrrrrr@{}}\toprule
&& KKE & SYRIZA & DIMAR & PASOK & ND & ANEL & GD\\
\midrule 
Myerson index && 0.05 & 0.18 & 0.22 & 0.18 & 0.32 & 0.05 & 0\\
\bottomrule
\end{tabular}
\caption{Myerson index for $50\%$ majority after May 06, 2012 elections.}
\label{myersonmay}
\end{table*}
The two leading parties, ND and SYRIZA, have an index close to their actual seat share. GD is not connected to any other party and hence, it has the role of a null player in the negotiations. On the other extreme, KKE is only connected -- as mentioned above only implicitly -- with the neighboring SYRIZA and hence, its power is less than its actual vote share. Yet, the most striking feature of the Myerson index is that it accurately captures DIMAR's influence in the negotiations and its crucial part in the formation of a coalitional government. Owing to its dense connections by its moderate stance, DIMAR has a Myerson-index of $22\%$ compared to its $6.33\%$ parliamentary seat share. Similar --yet less inflated due to less connections -- are the figures for PASOK. These observations can be compared with the \emph{a-priori} rankings induced by the non-selective Shapley-Shubik and Holler indices. 

\section{The June 12, 2012 Parliamentary election}\label{june}
\begin{figure}[!htb]
\begin{minipage}[b]{0.49\textwidth}\centering
\begin{tabular}{@{}llrrr@{}}
\toprule
&&\multicolumn{3}{c}{June 12, 2012} \\
\cmidrule{3-5}
\multicolumn{2}{@{}l}{Political Party} & \% Votes & \# Seats & \% Seats\\
\midrule
ND  && 29.00\% &129 & 43.00\% \\
SYRIZA && 26.89\% & 71 & 23.67\%\\
PASOK  && 12.28\% & 33 & 11.00\%\\
ANEL  && 7.51\% & 20 & 6.67\% \\
GD  && 6.92\% & 18 & 6.00\% \\
DIMAR && 6.25\% & 17 & 5.67\% \\
KKE  && 4.5\% & 12 & 4.00\%\\
\bottomrule
\end{tabular}
\captionof{table}{Votes share, Seats and Seat share.}
\label[table]{parliamentjune}
\end{minipage}
\hfill
\begin{minipage}[b]{0.49\textwidth}
\centering
\includegraphics[width=\linewidth]{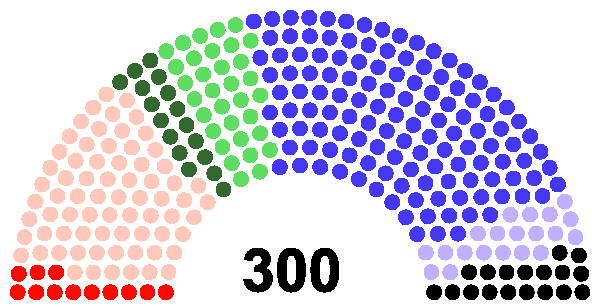}
\caption[]{Parliament, June 12, 2012.}
\label[figure]{parliamentjunefig}
\end{minipage}
\end{figure}

After the May deadlock, Greece headed for a second election, scheduled on June 12, 2012. The campaign took place under the same severe social and economic problems, extreme uncertainty and volatility as well as increasing external and domestic pressure to produce an outcome that would enable government formation, \cite{Va13,Di13} and \cite{Di15}. The results of the June 12, 2012 elections are presented in \Cref{parliamentjune} and \Cref{parliamentjunefig}. Again, the $50$ seats bonus has been awarded to ND and is incorporated in the figure below.
This time the negotiations were successful and led to the formation of a coalition government between ND (129 seats), PASOK (33 seats) and DIMAR (17 seats) that was enjoying a (seemingly) solid majority of 179 seats. It is worth mentioning that DIMAR was \emph{not} crucial anymore in this coalition. The configuration of the cabinet in August 2012, which did not correspond to the distribution of seats, \cite{Va13}, was composed of 13 ministers from ND, two nominated by PASOK and two nominated by DIMAR. In reality, the governing coalition proved problematic and after short time conflicts began, \cite{Ko15}. DIMAR withdrew from the government in June 2013. This left ND and PASOK with a slim majority of 162 seats, clearly short from the required 180 votes to win the upcoming Presidential election of December 2014 and remain in power.

\subsection{Measuring a priori power: the classic indices}\label{indicesjune}
\begin{figure}[!htb]
\centering
\includegraphics[]{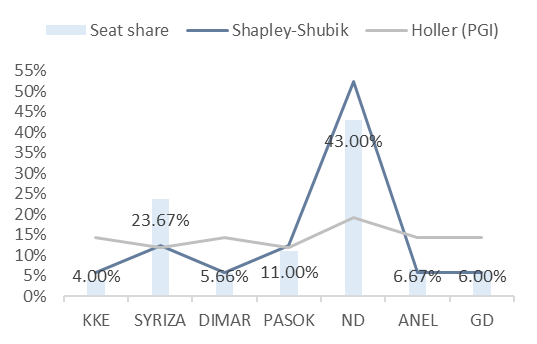}
\caption[]{Parliamentary seats share and main P-Power indices after June 12, 2012 elections.}
\label{powersharejune}
\end{figure}

The Greek Parliament after the June 12, 2012 elections can be described as the weighted voting game  $G=[151; 129,\ab 71,33,20,18,17,12]$. The Shapley-Shubik and Holler a priori power indices are given in \Cref{june50}. As in \Cref{indicesmay}, they are also depicted graphically, along with the parties' parliamentary seat shares in \Cref{powersharejune}. 
\begin{table*}[!htb]
\centering
\ra{1}
\begin{tabular}{@{}llrrrrrrr@{}}\toprule
&& \multicolumn{7}{c}{Political parties: June 12, 2012} \\
\cmidrule{3-9}
Power index & & KKE & SYRIZA & DIMAR & PASOK & \phantom{PS}ND & ANEL & \phantom{PS}GD\\
\midrule
Shapley-Shubik && 0.06 & 0.12 & 0.06 & 0.12 & 0.52 & 0.06 & 0.06\\
Holler && 0.14 & 0.12 & 0.14 & 0.12 & 0.19 & 0.14 & 0.14 \\
\bottomrule
\end{tabular}
\caption{Majority at 50+\% (151 MPs): a priori voting power in June 12, 2012.}
\label{june50}
\end{table*}
Compared to the May elections, our main findings are the following. The SSI now significantly deviates from the actual parliamentary seat share for both leading parties, ND and SYRIZA. Also, the SSI value of the governing partners ND, PASOK and DIMAR reflects the aforementioned difference between the assignment of ministers and their actual distribution of seats. This may provide a hint about the problematic nature of the governing coalition and the frictions between its members that surfaced soon after its formation. SYRIZA, the main opposition party, has an SSI value of $12\%$ despite a $23.67\%$ percentage of seats. This ranks SYRIZA equally with PASOK, and hence both parties have the same power to influence the negotiations, despite a $14\%$ difference in their vote shares. When it comes to minimal winning coalitions, the Holler index shows a remarkably uniform distribution of power with only a slight advantage for the leading ND. It is worth noting, that the governing coalition that eventually formed between ND, PASOK and DIMAR was not minimal. As in the May elections, SYRIZA is ranked second to last by the PGI. The low a-priori power of SYRIZA may be attributed to the electoral rule and the $50$ seat bonus that is given to the first party.

\subsection{Connections between parties and Myerson index}
Using the connectivity graph in \Cref{connect2012}, the Myerson index for the June 12, 2012 elections is given in \Cref{myerson2012b}. 
\begin{table*}[!htb]\centering
\ra{1}
\begin{tabular}{@{}llrrrrrrr@{}}\toprule
&& KKE & SYRIZA & DIMAR & PASOK & ND & ANEL & GD\\
\midrule 
Myerson Index && 0.03 & 0.12 & 0.12 & 0.28 & 0.42 & 0.03 & 0\\
\bottomrule
\end{tabular}
\caption{Myerson index for $50\%$ majority after June 12, 2012 elections.}
\label{myerson2012b}
\end{table*}
Again, the Myerson index provides some of the most interesting insights. ND, with a value close to its actual seat share and PASOK are identified as the two main political actors. Remarkably, SYRIZA's value has decreased from $0.18$ in May to $0.12$ in June, despite SYRIZA's gain of $20$ seats, whereas PASOK's index has surged to $0.28$ from $0.18$ in May, despite a loss of $8$ seats. While it is well known that the Myerson index is not monotonic in voting weights, this instance reflects the improvement in PASOK's strategic position due to its connection to the leading ND and the fact that their combined seat share now exceeds the required $50\%$ majority to form a government, cf. \Cref{parliamentjune}. In turn, this implies that SYRIZA, despite its increased seat share, has less power to influence the negotiations. Similarly, DIMAR's reduced value captures that DIMAR has forfeited its strategic advantage since it is no more crucial for the formation of a governing coaltion despite having retained its voting share and connections. Finally, the value of the three less connected parties, KKE, ANEL and GD has remained low in agreement with their marginal impact in the negotiations.

\section{The December 29, 2014 Presidential election}\label{presidential}
To more accurately assess the distribution of power between the elected parties under different institutional rules, we evaluate power indices at the $60\%$ majority which is relevant for the 2014 Presidential Election that was held by the same Parliament. We use the parties' seat shares after the June 12, 2012 elections and ignore subsequent strategic placements by parties or the independence movements by MPs. \par
As mentioned above, by December 2014, DIMAR had left the government and the remaining parties, ND and PASOK, had to ensure a parliamentary majority of 180 out of 300 MPs to elect the new President of the Republic and remain in power. This provided a doubtless test to the ruling coalition's stability and as \cite{Di15} correctly predicted, it enabled the opposition to lead the country to early elections in 2015. Eventually, the ND-PASOK candidate reached a total of 168 approval votes, 12 short of the required majority. The remaining 132 MPs concentrated in the disapproval camp which comprised the whole parliamentary teams of SYRIZA and ANEL, a part of the dissipating DIMAR and de facto the left and right extreme, anti-systemic parties, KKE and GD. Politically, SYRIZA's stance to align with the no-camp and block the Presidential election was rewarded by the electorate with a clear winning margin in the early elections of January 2015 that followed.   

\subsection{Measuring a priori power at \texorpdfstring{$60\%$}{k}}
Using the seat shares from the June 2012 elections, the Greek Presidential election of December 29, 2014  can be described as the weighted voting game $G=[180; 129,71,33,20,18,17,12]$. The a priori SSI and PGI are evaluated at the $60\%$ majority in the first part of \Cref{december2014}.
\begin{table}[!htb]
\centering
\begin{tabular}{@{}llrrrrrrr@{}}
\toprule
&& \multicolumn{7}{c}{Political parties: December 29, 2014} \\
\cmidrule{3-9}
Power index & & KKE & SYRIZA & DIMAR & PASOK & \phantom{PS}ND & ANEL & \phantom{PS}GD\\
\midrule
Shapley-Shubik && 0.01 & 0.16 & 0.03 & 0.08 & 0.63 & 0.05 & 0.03\\
Holler && 0.06 & 0.06 & 0.12 & 0.19 & 0.31 & 0.12 & 0.12 \\
\midrule
Coleman's to prevent && 0.02 & 0.36 & 0.06 & 0.23 & 1 & 0.11 & 0.11\\
Coleman's to initiate && 0.01 & 0.21 & 0.04 & 0.14 & 0.58 & 0.06 & 0.06\\
\bottomrule
\end{tabular}
\captionof{table}{Enhanced majority at 60\% (180 MPs): a priori voting power in December 29, 2014.}
\label{december2014}
\end{table}
The main finding of the SSI is that SYRIZA, the main opposition party, is now ranked second by a clear margin from the third party (in number of seats) PASOK. At the $50\%$ majority, SYRIZA and PASOK were tied at the second place, cf. \Cref{june50}. Moreover, at the $60\%$ quota, the smaller parties are ranked according to their seat share, cf. \Cref{parliamentjune}. The PGI attributes disproportionally high power to smaller parties and ranks SYRIZA in the last position together with KKE. \par
However, the main purpose of assessing the distribution of power at the $60\%$ quota, is to measure the governing coalition's power to pass an action and the power of the opposition to block this action. Thus, we focus on Coleman's indices of a party to \emph{prevent} (CP) and of a party to \emph{initiate} (CI) an action respectively, see \cite{Le02}. The results for the CP and CI indices are given in the second part of \Cref{december2014}. The CP index is the first to capture SYRIZA's uneven power to influence the outcome at the $60\%$ quota. Even more striking, is the power that the CP index attributes to the right extremist party, GD, to prevent action. Ideologically isolated from the other parties and a null player when it comes to forming winning coalitions, GD becomes an influential actor at preventing actions. Given the \emph{de facto} alignment of anti-systemic parties at the \enquote{no}-camp, this reveals an uneven advantage for the opposition to influence the outcome and block actions at the $60\%$ majority. A similar finding also holds for the ANEL who wield a disproportionally high power and have become a crucial political factor. Finally, the CI index attributes a combined power to ND and PASOK to initiate an action at the $60\%$ quota that is considerably less than $1$, indicating that a more broad cooperation is required to pass an action at this level.

\subsubsection{Strategic considerations}
The 2014 Presidential election was the first that took place in the presence of a high share of anti-systemic parties. Given the predisposition of such parties against any action in the Parliament, the above findings about the uneven power of the opposition to block actions at the enhanced majority raise concerns about the prevailing institutional rules. While a \enquote{yes}-vote indicates support to the proposed candidate and implicitly a vote of confidence to the government, the interpretation of a \enquote{no}-vote is more involved. It may range from disapproval of the specific candidate to distrust to the government. However, the \enquote{no}-vote also accommodates the anti-systemic voices of extremist parties that express their de facto disapproval of the process. In this way, ideologically opposing parties unavoidable align in the same camp and produce confusing electoral outcomes that do not represent the electorate's preferences. In the particular instance of the Presidential election, KKE, SYRIZA, ANEL and GD found themselves aligned in the same camp despite having very different and even competing platforms. The combined effect of this unavoidable misalignment is a distortion of the voting procedure and of the resulting outcome. As electoral rules change to accommodate systemic changes in the electorate's preferences, \cite{Or02}, the above findings provide an additional argument for the suggested revision of the Presidential electoral rule as part of a broader Constitutional revision, \cite{na18,he18}.

\section{Conclusions \& future research}\label{conclusions}
We employed power indices to revisit the twin Greek Parliamentary elections of May and June 2012 and the Presidential election of December 2014. We modeled the Greek Parliament as a weighted majority game and evaluated the Shapley-Shubik and Public Good (Holler) indices and when applicable the Myerson and Coleman's indices to assess the distribution of power among the elected political parties. The evaluation of the indices contributed to our understanding of the electoral results and, what is usually less considered, of the parties' electoral strategies. From a methodological point, our findings provide empirical evidence on the effectiveness of non-selective indices such as Shapley-Shubik and Public Good in the densely connected, multi-party setting of the present case-study. Beyond their traditional application in committee settings, this performance indicates that power indices may prove useful in similar studies on the current European Parliaments of increasing diversity and connectivity among political parties.\par
In the current political context, our findings shed light on the influence of electoral rules in the distribution of power among elected parties and provide an input to the heated political debate in Greece about the need to reform both the Parliamentary and Presidential electoral rule, \cite{he18}. \cite{Te14} argue in favor of a simple proportionality system to mitigate the pressure for strategic voting, a thesis also adopted by the incumbent government, \cite{ek16}. Such proposals naturally raise the question whether the new electoral rules will adequately capture public preferences or not and require closer scrutiny, \cite{Ke12}. In a different direction, \cite{Ko15} and \cite{Ts16} observe that Greek voters become increasingly favorable towards coalitional governments and highlight the need to study the concentration or fragmentation of political power in the evolving party system. All the above attest to a rapidly changing political ecosystem in Greece that forms an attractive research topic for the mathematical, social and political sciences.

%\section*{Acknowledgements}\label{acknow}
%The present study would not have been possible without the freely available software \href{http://homepages.warwick.ac.uk/~ecaae/}{ipdirect} due to Denis and Robert Leech and \href{https://alex.uniupo.it/software-english}{AL.EX4} due to Marie-Edith Bissey, Vito Fragnelli and Guido Ortona. We would also like to thank Manfred Holler, Guido Ortona and Josep Freixas for providing valuable feedback on earlier versions of this paper and the editor Stefan Napel and an anonymous referee for valuable suggestions to improve our work.

\section*{Conflict of Interest -- Acknowledgments}
On behalf of all authors, the corresponding author states that there is no conflict of interest. Although we would like to thank several persons for the completion of this work, we avoid to do so at this iteration to avoid any potential conflict with prospective referees. 

\bibliographystyle{plainnat}	
\bibliography{2012greekbib}
\end{document}